\documentclass{article}

\usepackage{graphicx}

\begin{document}
\title{Stability of circular orbits of spinning particles in Schwarzschild-like space-times}
\author{Morteza Mohseni\thanks{E-mail address:m-mohseni@pnu.ac.ir}\\
\small Physics Department, Payame Noor University, 19395-4697
Tehran, Iran}

\maketitle
\begin{abstract}
Circular orbits of spinning test particles and their stability in Schwarzschild-like backgrounds are investigated. For these 
space-times the equations of motion admit solutions representing circular orbits with particles spins being constant and normal to the plane of
orbits. For the de Sitter background the orbits are always stable with particle velocity and momentum being co-linear along them. The world-line
deviation equations for particles of the same spin-to-mass ratios are solved and the resulting deviation vectors are used to study the stability
of orbits. It is shown that the orbits are stable against radial perturbations. The general criterion for stability against normal
perturbations is obtained. Explicit calculations are performed in the case of the Schwarzschild space-time leading to the conclusion that the
orbits are stable. 

keywords: Stability of orbits, Spinning particles

\end{abstract}
\section{Introduction}
Orbits of particles in curved space-times have long been a topic of interest in physics and astronomy. In relativistic theories of motion,
orbits of particles around astronomical objects have been studied extensively in the past and have resulted in a plethora
of phenomena ranging from the perihelion precession and bending of light to more recent ones like the so-called clock effect \cite{cohen}.
Nearly circular orbits have had a especial role in this regard. Their application in theories dealing with accretion discs 
\cite{pringle} (see also \cite{kato} and references therein) is an example showing their importance. Some of the objects moving on such orbits 
may have some internal angular momentum. In trying to model the motion of such rotating objects treated as test particles, one is lead to the 
Mathisson-Papapetrou-Dixon (MPD) equations \cite{mathi,papa,tulc,dixon}. According to these equations, in general the trajectories of a spinning 
particle moving in a curved background differ from the geodesics of the background space-time as a result of a coupling between the particle 
spin and the space-time curvature and furthermore, its momentum and velocity are not in general parallel.

Being a system of highly coupled differential equations, the MPD equations are usually difficult to solve and only a few analytical solutions to
them are known, e.g. the one introduced in \cite{tod} in which the motion in the equatorial plane of a Kerr black hole with a fixed particle
spin was investigated. Following the ideas presented in the latter reference, solutions describing circular or nearly circular orbits in various 
black hole space-times have been found. Different aspects of such orbits have been studied in the case of Lense-Thirring space-time in 
\cite{ryab1}, in the case of Schwarzschild space-time in \cite{svir,ryab2,bini,geralico2,plya} (also in \cite{bakh} for non-circular orbits), in 
the case of Kerr black hole in \cite{seme,hart1,hart2,faru,geralico,felice,singh,kyr,singh2}, in the case of Reissner-Nordstr\"{o}m 
black hole in \cite{bini2}, and in the case of Weyl space-time in \cite{binib}. An alternative framework has been used in \cite{van,rit} and 
\cite{gib} to study orbits of spinning particles in Schwarzschild and Kerr-Newman space-times respectively.

An interesting question is whether these orbits are stable. For the case of (spinless) test particles this question has been considered
in \cite{shir,fuchs} where deviation of circular geodesics in static spherically symmetric space-times was studied. The aim of the present work 
is to investigate this (in)stability for orbits of spinning particles. The technique we use in this regard is to study the growth of 
perturbations to a given orbit. These perturbations are essentially deviations from a fiducial orbit to adjacent ones. Unbounded growth of a 
perturbation signals instability of the orbit. A recent application of this technique may be found in \cite{rosa} in which the geodesic 
deviation equation is used to study the stability of circular orbits of test particles in Einstein-Gauss-Bonnet gravity. In fact as it has been 
argued in \cite{wu} the deviation vectors may be used to define Lyapunov indices in general relativity for either geodesic or non-geodesics 
flows. The method of Lyapunov indices has been applied to study the stability of circular or non-circular orbits of spinning particles in 
\cite{ryab1,svir,ryab2,bakh}. In particular in \cite{ryab2} it was shown that using different supplementary conditions may result in different 
stability behaviour of spinning particles orbiting the Schwarzchild black hole. The problem of chaotic motion of spinning particles in 
Schwarzschild space-time has also been studied in \cite{suzuki} but in that work the deviation equations were not used and a numerical scheme 
based on a time series method was used instead. Chaotic motion of spinning particles around the Schwarzschild black hole has also been studied 
in \cite{sano} from a different perspective. The case of Kerr balck hole has been considered in \cite{suz}.
  
To study perturbations of spinning particles orbits, the geodesic deviation equation is no longer the appropriate equation, instead we should 
use its generalization as presented in \cite{nieto} or \cite{mohseni}. A similar set of equations have been obtained in \cite{sepangi} for the
case of charged spinning particles. In this work we deploy the machinery developed in \cite{mohseni} to obtain the deviations. 

As it has been argued in \cite{muller} the test particle approach leading to equations like the MPD equations, is valid only when the particles
spins are small compared to the characteristic length of the background they are moving in. Thus throughout this work we assume the particles
specific spins (i.e., their spin-to-mass ratios) to be small compared with the mass of the black hole.

The paper is organized as follows. In the next section we give a brief review of the MPD equations and show how a solution describing a circular
orbit can be found in a rather general Schwarzschild-like space-time. Although circular orbits in particular spherically symmetric backgrounds 
have been studied in detail in some of the above mentioned references, that section serves both as a rather unified manner review of the main 
results essential for the subsequent sections and to fix the notation. The latter is of importance because a variety of notations have been used 
in the literature. These include the way in which the gauge is fixed and the particle spin tensor is related to the spin vector. The case of de 
Sitter space-time is simpler and is studied in a separate section as a warm up. Then we solve the world-line deviation equations for circular 
orbits and use them to study the behaviour of the perturbations. Even though our focus of attention is on orbits in Schwarzschild space-time, we 
obtain some general criteria under which the orbits are (un)stable in a general Schwarzschild-like space-time. The paper is concluded with a 
discussion of the results.
\section{Circular Orbits}
The basic equations governing the motion of a spinning test particle are \cite{dixon}
\begin{eqnarray}
\frac{Dp^\mu}{D\tau}&=&-\frac{1}{2}R^\mu{}_{\nu\alpha\beta}v^\nu s^{\alpha\beta},\label{eg0}\\
\frac{Ds^{\mu\nu}}{D\tau}&=&p^\mu v^\nu-p^\nu v^\mu,\label{eg1}\\
p_\mu s^{\mu\nu}&=&0\label{eg2}
\end{eqnarray}
in which $v^\mu=\frac{dx^\mu(\tau)}{d\tau}, p^\mu, s^{\mu\nu}$ are the particle velocity, momentum, and spin respectively, $\frac{D}
{D\tau}=v^\alpha\nabla_\alpha$ with $\nabla_\mu$ being the covariant derivative, and
$${R^\mu}_{\nu\alpha\beta}=\partial_\alpha{\Gamma^\mu}_{\beta\nu}-\partial_\beta{\Gamma^\mu}_{\alpha\nu}+{\Gamma^\mu}_{\alpha\delta}
{\Gamma^\delta}_{\beta\nu}-{\Gamma^\mu}_{\beta\delta}{\Gamma^\delta}_{\alpha\nu}.$$ represents the space-time curvature tensor. The particle
spin is related to its spin vector by \cite{tod}
\begin{equation}\label{eg3}
s^{\mu\nu}=\frac{1}{m\sqrt{-g}}\epsilon^{\mu\nu\alpha\beta}p_\alpha s_\beta
\end{equation}
with $\epsilon^{\mu\nu\alpha\beta}$ being the alternating symbol. In this work we adopt ${\epsilon}^{tr\theta\phi}=+1$. It should be noted that
a different defining equation is sometimes used in this regard (see e.g. equation A1 of \cite{suzuki}). As a consequence of these equations one
can show that the particle mass and spin are constant
\begin{eqnarray}
p_\mu p^\mu&=&-m^2=const.,\label{eg0a}\\
s_\mu s^\mu&=&s^2=const.\label{eg0b}
\end{eqnarray}
We choose the following gauge fixing relation \cite{dixon}
\begin{equation}\label{eg7}
p_\mu v^\mu=-m
\end{equation}
in which the particle instantaneous zero-momentum and zero-velocity frames are simultaneous \cite{ehler}.
This is the gauge in which the velocity-momentum relation \cite{tod,ehler}
\begin{equation}\label{equate}
-\frac{m^2}{v_\alpha p^\alpha}v^\mu=p^\mu+\frac{2s^{\mu\nu}R_{\nu\rho\alpha\beta}
s^{\alpha\beta}}{4m^2+s^{\mu\nu}R_{\mu\nu\alpha\beta}s^{\alpha\beta}}p^\rho
\end{equation}
which is itself a consequence of the equations of motion, takes the simplest form.

We consider the motion of a spinning test particle in a Schwarzschild-like background space-time described in the coordinate system 
$(t,r,\theta,\phi)$ by the line element
\begin{equation}\label{eg4}
ds^2=-f(r)dt^2+\frac{1}{f(r)}dr^2+r^2(d\theta^2+\sin^2\theta d\phi^2)
\end{equation}
in which the explicit form of $f(r)$ is subject to the Einstein field equations. This metric corresponds to Schwarzschild space-time for 
$f(r)=1-\frac{2M}{r}$, to Reissner-Nordstr\"{o}m space-time for $f(r)=1-\frac{2M}{r}+\frac{Q^2}{r^2}$, to de Sitter space-time 
$f(r)=1+\frac{\Lambda}{3}r^2$, to Schwarzschild-de Sitter space-time for $f(r)=1-\frac{2M}{r}+\frac{\Lambda}{3}r^2$, and to 
Reissner-Nordstr\"{o}m- de Sitter space-time for $f(r)=1-\frac{2M}{r}+\frac{Q^2}{r^2}+\frac{\Lambda}{3}r^2$.

Inspired by the results of \cite{tod}, we consider a spinning particle moving on a circular orbit $r=a$ in the equatorial plane 
$\theta=\frac{\pi}{2}$ with a velocity $v^\mu=(v^t,0,0,v^\phi)$. We take the particle spin vector perpendicular to this plane. Thus we set
\begin{equation}\label{eg5a}
s^\mu=\frac{1}{r}(0,0,s,0)
\end{equation}
and look for a consistent solution
\begin{equation}\label{eg5b}
p^\mu(\tau)=(p^t(\tau),p^r(\tau),0,p^\phi(\tau))
\end{equation}
of the equations of motion (\ref{eg0}) and (\ref{eg1}). Note that Eqs. (\ref{eg5a}) and (\ref{eg5b}) are consistent with Eq. (\ref{eg2}) by 
construction. Now making use of Eq. (\ref{eg3}), we can insert this ansatz into Eq. (\ref{eg0}), to obtain
\begin{eqnarray}
\frac{dp^t}{d\tau}&=&-\frac{f^\prime}{2f}v^tp^r+\frac{{\hat s}f^\prime}{2f}v^\phi p^r,\label{ep1}\\
\frac{dp^r}{d\tau}&=&f\left(-\frac{f^\prime}{2}v^tp^t+av^\phi p^\phi+\frac{{\hat s}}{2}
(-f^{\prime\prime}av^tp^\phi+f^\prime v^\phi p^t)\right),\label{ep2}\\
\frac{dp^\phi}{d\tau}&=&-\frac{1}{a}v^\phi p^r-\frac{{\hat s}f^\prime}{2a^2}v^tp^r,\label{ep3}
\end{eqnarray}
where $f=f(a)$, $\displaystyle f^\prime=\left.\frac{df(r)}{dr}\right|_a,$ and $\displaystyle{\hat s}=\frac{s}{m}$. Similarly, equation
(\ref{eg1}) leads to
\begin{eqnarray}
\frac{{\hat s}f}{a}\frac{dp^t}{d\tau}&=&-\frac{{\hat s}f^\prime}{2a}v^tp^r+v^\phi p^r ,\label{ep1a}\\
\frac{{\hat s}}{fa}\frac{dp^r}{d\tau}&=&-\frac{{\hat s}f^\prime}{2a}v^tp^t+{\hat s}v^\phi p^\phi+v^\phi p^t-v^tp^\phi,\label{ep2a}\\
{\hat s}a\frac{dp^\phi}{d\tau}&=&-{\hat s}v^\phi p^r+v^tp^r.\label{ep3a}
\end{eqnarray}
Solving Eqs. (\ref{ep1}),(\ref{ep3}),(\ref{ep1a}), and (\ref{ep3a}) for the components of $p^\mu$ lead to
\begin{eqnarray}
p^t&=&const.,\label{eg8a}\\
p^r&=&0,\label{eg8b}\\
p^\phi&=&const.\label{eg8c}
\end{eqnarray}
Now the remaining two equations, (\ref{ep2}) and (\ref{ep2a}), together with Eqs. (\ref{eg0a}) and (\ref{eg7}) reduce to the following system of 
algebraic equations
\begin{eqnarray}
f{p^t}^2-a^2{p^\phi}^2&=&m^2\label{ex1},\\
fv^tp^t-a^2v^\phi p^\phi&=&m,\label{ex2}\\
\frac{f^\prime}{2}v^tp^t-av^\phi p^\phi&=&\frac{{\hat s}f^\prime }{2}v^\phi p^t-\frac{{\hat s}f^{\prime\prime}a}{2}v^tp^\phi,\label{ex3}\\
\frac{{\hat s}f^\prime}{2a}v^tp^t-{\hat s}v^\phi p^\phi&=&v^\phi p^t-v^tp^\phi\label{ex4}
\end{eqnarray}
from which we can obtain the non-vanishing components of $v^\mu$ and $p^\mu$.
\section{Motion in de Sitter space-time}
To proceed further, we first consider the simplest case where $v^\mu$ and $p^\mu$ are parallel, i.e. $p^\mu=mv^\mu$ but ${\hat s}\neq 0$.
Taking this into account, the above system of algebraic equations are consistent only for an $f(r)$ of the form of $f(r)=1+Ar^2$, $A$ being a
constant. This corresponds to the well-known de Sitter space-time. Thus we set
\begin{eqnarray}
f(r)=1+\frac{\Lambda}{3}r^2.
\end{eqnarray}
This restriction means that there is no solution describing a circular orbit with velocity and momentum being parallel in other 
Schwarzschild-like space-times. From the fact that de Sitter space-time is homogeneous we already expected $v^\mu$ and $p^\mu$ to be parallel.
This is easily confirmed by the above equations resulting in
\begin{equation}\label{eg9}
p^\mu=mv^\mu=m\left(1,0,0,\pm\sqrt{\frac{\Lambda}{3}}\right).
\end{equation}
The angular velocity of particle is independent of the orbit it moves on.

We now aim to investigate the stability of these orbits. To this end, we consider the behaviour of perturbations in different directions.
Unbounded growth of a perturbation is a signature of instability. A suitable framework to study stabilities is the world-line deviation
equations developed in \cite{mohseni}. However, in this case the momentum and velocity four-vectors are co-linear and the equations of motion
reduce to the geodesic equation. In fact for maximally symmetric space-times we have
\begin{equation}
R_{\mu\nu\alpha\beta}=C(g_{\mu\alpha}g_{\nu\beta}-g_{\mu\beta}g_{\nu\alpha})
\end{equation}
where $C$ is constant. Inserting this into Eq. (\ref{eg0}) returns $$\frac{Dp^\mu}{D\tau}=-Cv^\nu {s^\mu}_\nu.$$
On the other hand by combining the above relation with Eq. (\ref{equate}) and making use of Eq. (\ref{eg2}) we arrive at the previously 
mentioned result $p^\mu=mv^\mu$. Thus $$\frac{Dp^\mu}{D\tau}=-\frac{C}{m}p^\nu{s^\mu}_\nu=0,$$ i.e. particles move along geodesics and their 
momentum and spin are parallel. Hence we can deploy the well-known geodesic deviation equation instead of the world-line deviation equations. 
It reads
\begin{equation}\label{es1}
\frac{D^2n^\mu}{D\tau^2}=-R^\mu{}_{\alpha\nu\beta}n^\nu v^\alpha v^\beta
\end{equation}
where $n^\mu=\frac{dx^\mu}{d\lambda}$ is the deviation between adjacent trajectories $x^\mu, x^\mu+\lambda n^\mu$.
Applying this equation to the above circular orbits results in
\begin{eqnarray}
\frac{d^2 n^t}{d\tau^2}&=&-\frac{f^\prime}{f}v^t\frac{d n^r}{d\tau},\label{es2}\\
\frac{d^2 n^r}{d\tau^2}&=&-ff^\prime v^t\frac{d n^t}{d\tau}+2af v^\phi\frac{d n^\phi}{d\tau}-kn^r,\label{es3}\\
\frac{d^2 n^\theta}{d\tau^2}&=&0,\label{es4}\\
\frac{d^2 n^\phi}{d\tau^2}&=&-\frac{2}{a}v^\phi\frac{d n^r}{d\tau}\label{es5}
\end{eqnarray}
where $$k=\frac{1}{2}({f^\prime}^2+ff^{\prime\prime})(v^t)^2-(af^\prime+f)(v^\phi)^2.$$ 
By integrating the first, the third, and the last equation in the above system of equations we obtain
\begin{eqnarray}
\frac{d n^t}{d\tau}&=&-\frac{f^\prime}{f}v^tn^r+C_1,\label{es6}\\
n^\theta&=&C_2t+C_3\label{es7},\\
\frac{d n^\phi}{d\tau}&=&-\frac{2}{a}v^\phi n^r+C_4\label{es8}
\end{eqnarray}
in which $C_1, C_2, C_3, C_4$ are constants. The second equation above shows that the circular orbits considered here are stable against normal 
(i.e., in the $\theta$-direction) perturbations. To see the effect of radial perturbations, we first insert the values of $f,v^t,v^\phi$ into 
the expression for $k$ to get $k=\frac{4\Lambda}{3}$. We then insert this back into Eq. (\ref{es3}) and solve it for $n^r$ with the aid of 
Eqs. (\ref{es6}) and (\ref{es7}) to obtain
\begin{equation}\label{es9}
n^r=N\cos(2v^\phi\tau)+l
\end{equation}
in which $N,l$ are constants. This leads us to the conclusion that the orbits are stable against radial perturbations.
\section{Motion in other Schwarzschild-like space-times}
Relaxing the simplifying assumption of the previous section, we now seek general solutions $p^\mu,v^\mu$ to Eqs. (\ref{ex1})-(\ref{ex4}).
In general these equations admit two sets of solutions which, roughly speaking, correspond to two senses of rotation with respect to a given 
spin direction. Labelling the solutions by $(p^t_+,p^\phi_+,v^t_+,v^\phi_+)$ and $(p^t_-,p^\phi_-,v^t_-,v^\phi_-)$ respectively, we have
\begin{equation}\label{sol2}
p^t_\pm=\frac{2(2-{\hat s}^2f^{\prime\prime})}{\sqrt{b_\pm}},
\end{equation}
\begin{equation}\label{sol3}
p^\phi_\pm=\frac{2({\hat s}^2f^{\prime\prime}-2)({\hat s}^3{f^\prime}^2+{\hat s}a^2f^{\prime\prime}-3a{\hat s}f^\prime\pm ah)}{a\sqrt{b_\pm}
({\hat s}^2af^{\prime\prime}+{\hat s}^2f^\prime-4a\pm{\hat s}h)}
\end{equation}
\begin{equation}\label{sol1}
v^\phi_\pm=\frac{-{\hat s}f^\prime+{\hat s}af^{\prime\prime}\pm h}{2(2a-{\hat s}^2f^\prime)}v^t,
\end{equation}
\begin{equation}\label{sol4}
v^t_\pm=\frac{2(2a-{\hat s}^2f^\prime)}{2f(2a-{\hat s}^2f^\prime)p^t-a^2(-{\hat s}f^\prime+{\hat s}af^{\prime\prime}\pm h)p^\phi},
\end{equation}
where
\begin{eqnarray*}
h=\sqrt{-3{f^\prime}^2{\hat s}^2-6a{f^\prime}{\hat s}^2f^{\prime\prime}+{\hat s}^2a^2{f^{\prime\prime}}^2+8{f^\prime}a+2{f^\prime}^2{\hat 
s}^4f^{\prime\prime}},
\end{eqnarray*}
\begin{eqnarray*}
b_\pm&=&-2{\hat s}^2a^2{f^{\prime\prime}}^2+8a{f^\prime}{\hat s}^2f^{\prime\prime}-8{f^\prime}a+2{f^\prime}^2{\hat s}^2-2{ f^\prime}^2{\hat 
s}^4f^{\prime\prime}\\&&-16f{\hat s}^2f^{\prime\prime}+4f{\hat s}^4{f^{\prime\prime}}^2+16f\pm 2{\hat s}(f^\prime-af^{\prime\prime})h
\end{eqnarray*}
An interesting property of these solutions is that under ${\hat s}\rightarrow-{\hat s}$ the absolute values of $v^\phi$ and $p^\phi$ change
in general. This is due to the spin-dependent character of the force exerted on the particles. Also note that the $+$ and $-$ solutions do not
represent physically distinct situations. This can be easily seen from the above expressions in which ${\hat s}\leftrightarrow -{\hat s}$ 
interchanges $+$ and $-$ solutions (up to an overall minus sign). In fact only two of the four possible situations $\{{\hat s}=\pm, 
sense\,of\,rotation=clokwise, counter-clockwise\}$ are independent because there is no preferred $\mbox{up}$ or $\mbox{down}$ direction in a 
spherically symmetric space. 

\begin{figure}[h]
\begin{center}
\includegraphics[width=8cm]{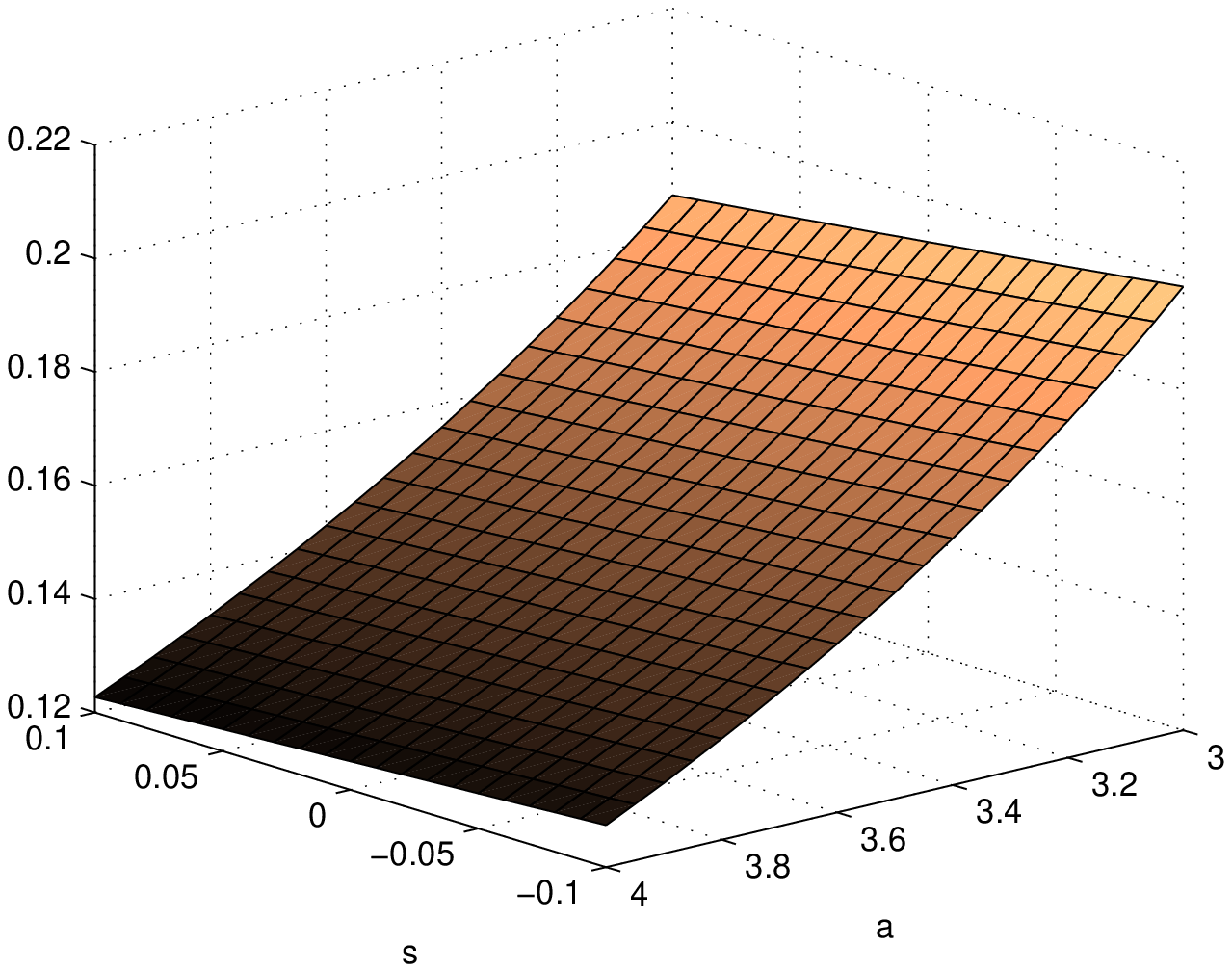}\includegraphics[width=7cm]{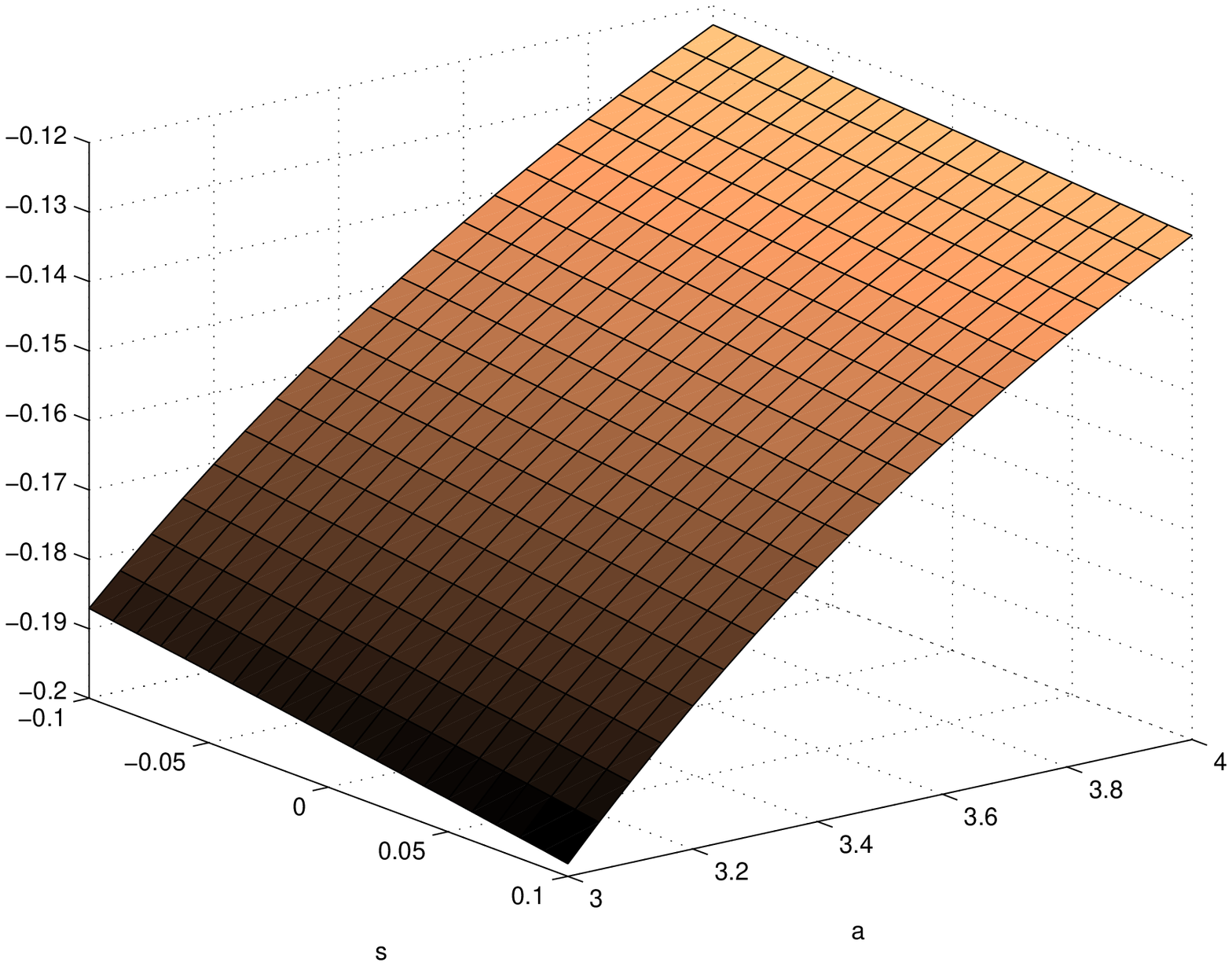}
\end{center}
\caption{The angular velocity of particles $\frac{v^\phi_+}{v^t_+}$ (left) and $\frac{v^\phi_-}{v^t_-}$ (right) in terms of $a$ and ${\hat 
s}$.}\label{depi1} 
\end{figure}

To get more physical insight into the nature of the above solutions we consider the case of Schwarzschild space-time. The functions 
${v^\phi}/{v^t}$ are depicted in Fig. \ref{depi1} in which the range of ${\hat s}$ is chosen as $|{\hat s}|<0.1$ to conform with the  
requirement ${\hat s}\ll M$ and the range of $a$ is taken as $a>3$ since for $2<a<3$, $v^t$ becomes imaginary, i.e. there is no circular orbit 
for this range. The case $a=3$ needs special care and is discussed later.

\begin{figure}[h]
\begin{center}
\includegraphics[width=7cm]{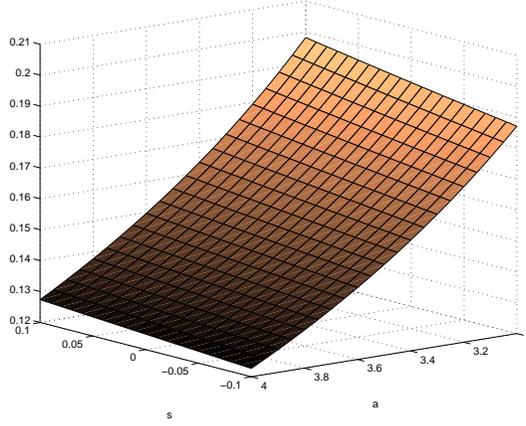}
\end{center}
\caption{The function $|\frac{v^\phi_-}{v^t_-}|$ in terms of $a$ and ${\hat s}$.}\label{depi7}
\end{figure}

The asymmetry ${\hat s}\leftrightarrow -{\hat s}$ is obvious in these graphs. They also show that for large values of $a$ the two solutions 
coincides. To ease comparison, the absolute value of $|{v^\phi_-}/{v^t_-}|$ is depicted in Fig. \ref{depi7}. It coincides with the  
left-hand side plot in Fig. \ref{depi1} upon taking ${\hat s}\to{-\hat s}$.

\begin{figure}
\begin{center}
\includegraphics[width=7cm]{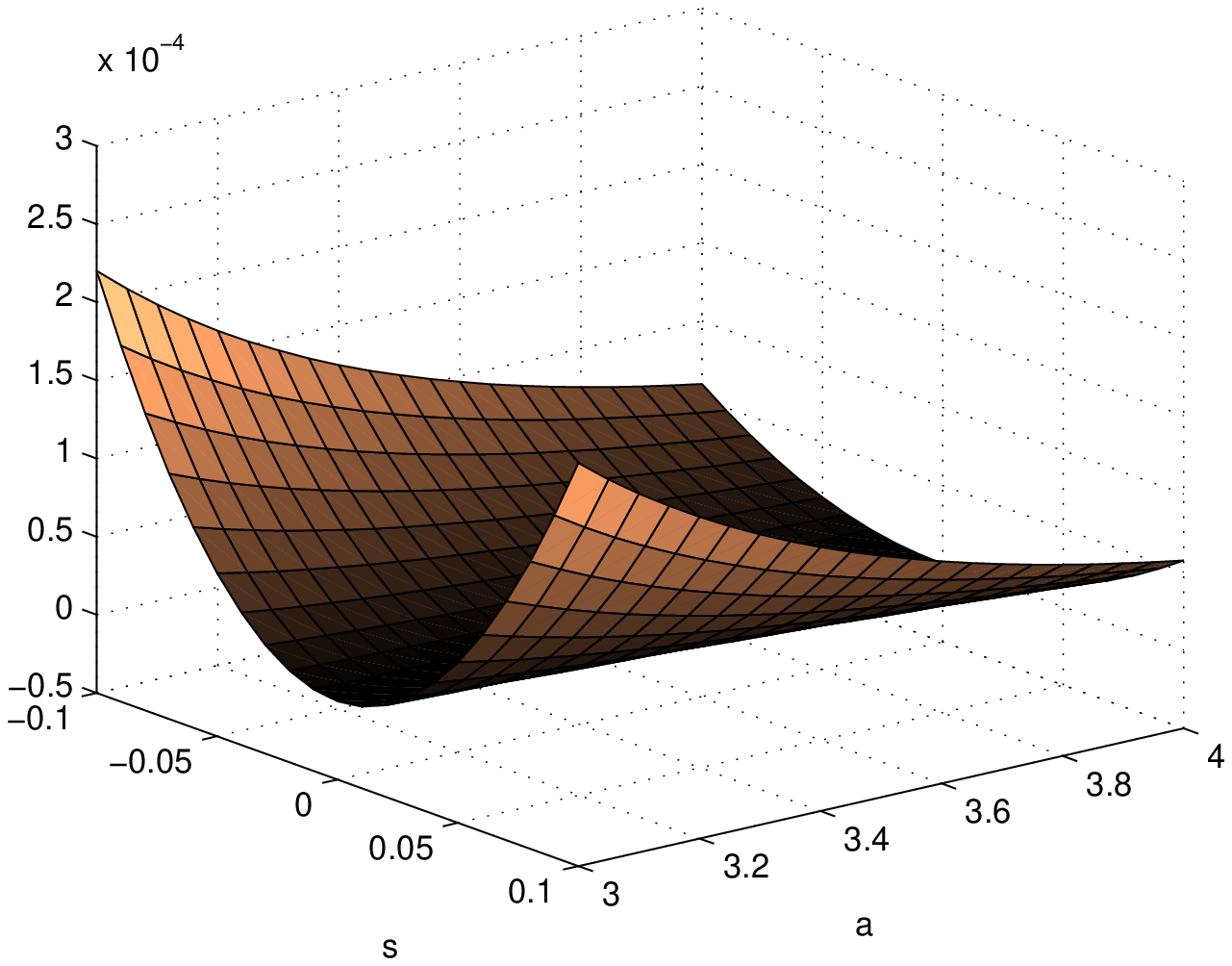}\includegraphics[width=7cm]{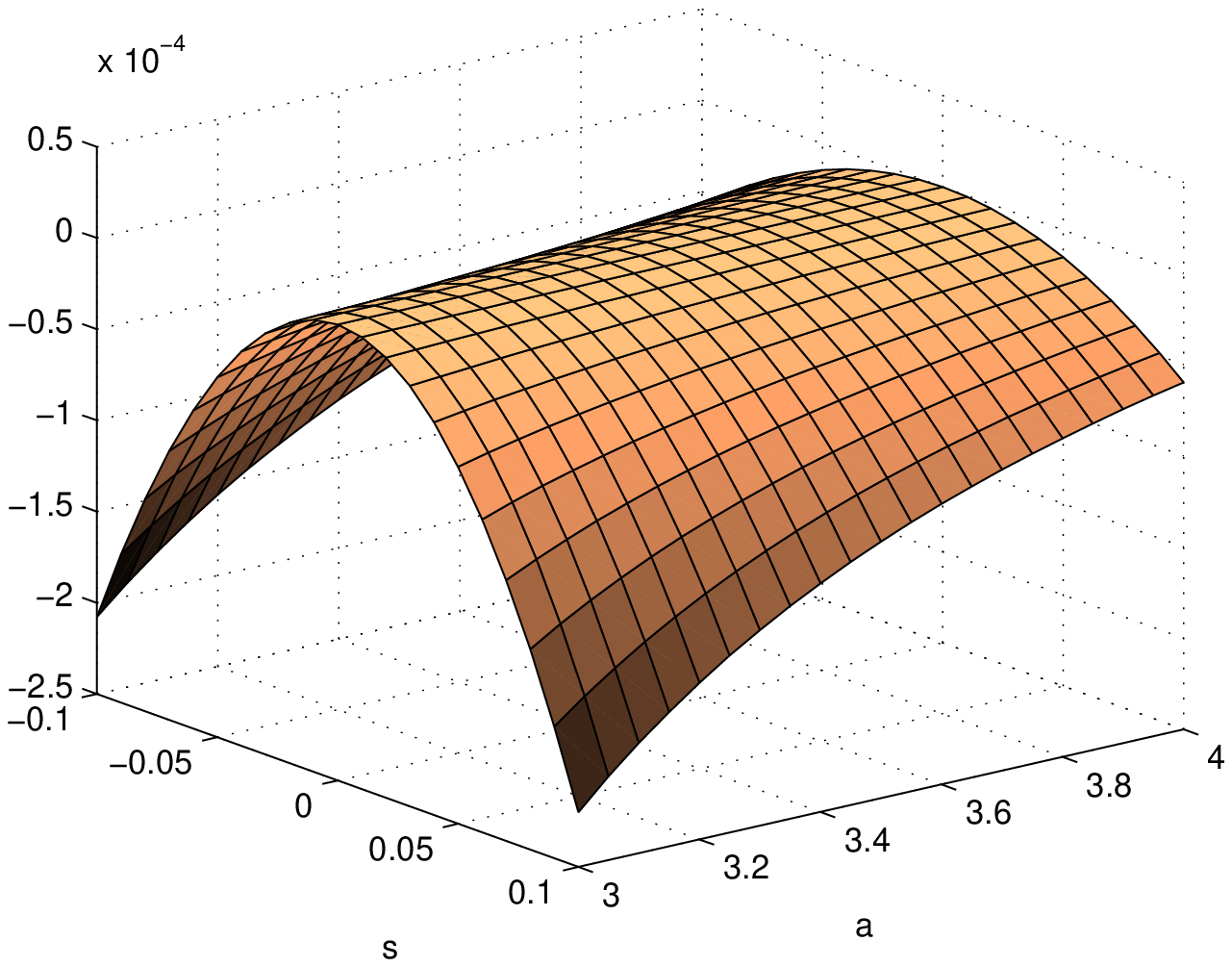}
\end{center} 
\caption{The non-co-linearity $\frac{v^\phi_+}{v^t_+}-\frac{p^\phi_+}{p^t_+}$ (left) and $\frac{v^\phi_-}{v^t_-}-\frac{p^\phi_-}{p^t_-}$ (right) 
in terms of $a$ and ${\hat s}$. }\label{depi2}
\end{figure}

We have also depicted the functions ${v^\phi_\pm}/{v^t_\pm}-{p^\phi_\pm}/{p^t_\pm}$ in Fig. \ref{depi2}. It shows that for larger values of 
$a$ the difference between ${p^\phi}/{p^t}$ and ${v^\phi}/{v^t}$ diminishes. This is because a larger $r=a$ results in a value of $f$ closer to 
unity and hence a smaller curvature which in turn results in a smaller spin-curvature coupling.  

To investigate the special case of $a=3$ mentioned earlier, we first consider the case of a particle without spin. For ${\hat s}=0$, 
Eq. (\ref{ex4}) is satisfied identically, and the remaining Eqs. (\ref{ex1})-(\ref{ex3}) result in 
\begin{equation}\label{ey1a}
v^t=\sqrt{\frac{2}{2f-af^\prime}}
\end{equation}
which for the Schwarzschild space-time implies $a>3$. When ${\hat s}\neq 0$, Eqs. (\ref{ex1})-(\ref{ex4}) do admit a solution given by 
\begin{equation}\label{ey22}
v^\phi_\pm=\frac{-27{\hat s}\pm l}{18(27-{\hat s}^2)}v^t,
\end{equation}  
\begin{eqnarray}\label{ey5}
p^t_\pm=\frac{\sqrt{6{\hat s}(27-{\hat s}^2)(2{\hat s}^3-135{\hat s}\pm 3l)}(-4{\hat s}^3+27{\hat s}\pm 3l)}{2{\hat s}(27-{\hat s}^2)(9{\hat 
s}^2\mp{\hat s}l+486)},
\end{eqnarray}
\begin{equation}\label{ey8}
p^\phi_\pm=\frac{\sqrt{6{\hat s}(27-{\hat s}^2)(2{\hat s}^3-135{\hat s}\pm 3l)}}{18{\hat s}( 27-{\hat s}^2)}
\end{equation}
where $l=\sqrt{-24{\hat s}^4+1053{\hat s}^2+8748}$ and $\pm$ in Eqs. (\ref{ey22}) and (\ref{ey8}) are chosen according to the sign of 
${\hat s}$. 

Two further remarks are in order. First, an ansatz corresponding to particles moving on circular orbits with the spin vector making an arbitrary
angle with the plane of the orbit, say $s^\mu\propto(0,s,0,0)$,  is not in general compatible with the equations of motion. Second, since the
Schwarzschild-like space-times are spherically symmetric there is no preferred equatorial plane and hence similar circular orbits on other
planes are also allowed.
\section{Stability of orbits}
In general, orbits of spinning particles are not geodesics of the background they are moving in and we can not use the geodesic deviation
equation.  In this work we use the world-line deviation equations \cite{mohseni}
\begin{eqnarray}
\frac{D j^\mu}{D\tau}&=&-\frac{1}{2}\nabla_\kappa {R^\mu}_{\nu\alpha\beta}n^\kappa v^\nu
s^{\alpha\beta}-\frac{1}{2}{R^\mu}_{\nu\alpha\beta}v^\nu J^{\alpha\beta}
\nonumber\\&&-\frac{1}{2}{R^\mu}_{\nu\alpha\beta}\frac{Dn^\nu}{D\tau}s^{\alpha\beta}
-{R^\mu}_{\beta\alpha\kappa}v^\kappa n^\alpha p^\beta,\label{e5}\\
\frac{D J^{\mu\nu}}{D\tau}&=&s^{\kappa[\mu}{R^{\nu]}}_{\kappa\alpha\beta}n^\alpha
v^\beta+p^{[\mu}\frac{Dn^{\nu]}}{D\tau}+j^{[\mu}v^{\nu]},\label{e4}\\
s_{\mu\nu}j^\nu&=&-J_{\mu\nu}p^\nu\label{e6}
\end{eqnarray}
where $j^\mu=n^\alpha\nabla_\alpha p^\mu$ and $J^{\mu\nu}=n^\alpha\nabla_\alpha s^{\mu\nu}$. The dimensions of various quantities used in these 
equations are shown in Table \ref{tab}.
\begin{table}[h]
\centering
\caption{The dimensions of various quantities used in world-line deviation equations. Here $L$ abbreviates length, $T=t,r$, and $\psi=\phi,
\theta$}\label{tab}
\begin{tabular}{|c|c|c|c|c|c|c|c|c|c|c|c|c|}\hline
Quantity & ${\hat s}$ & $p^T$ & $p^\psi$ & $n^T$ & $n^\psi$ & $j^T$ & $j^\psi$ & $J^{TT}$ & $J^{T\psi}$ & $J^{\psi\psi}$ \\ 
\hline Dimension & $L$ & $L$ & $L^0$ & $L$ & $L^0$ & $L$ & $L^0$ & $L^2$ & $L$ & $L^0$ \\ \hline
\end{tabular}
\end{table}

The above equations are the generalization of geodesic deviation equation to world-line deviations of spinning particles of the same specific 
spins. In fact setting $s^{\mu\nu}=0$ in Eq. (\ref{e4}) immediately results in $$p^\mu=b\frac{Dn^\mu}{D\tau}$$ in which $b$ is a constant. 
Inserting this back into Eq. (\ref{e5}) we arrive at Eq. (\ref{es1}), the geodesic deviation equation. It is interesting to note that, 
according to Eq. (\ref{e5}), even if we neglect the direct effect of spin, i.e. the first three terms in the right hand side of that 
equation, the particles spin still affect the dynamics through $p^\mu$ in the last term which in turn depends on spin. The following additional 
equations
\begin{eqnarray}
p_\mu j^\mu&=&0\label{ey1},\\
p_\mu J^{\mu\nu}&=&-j_\mu s^{\mu\nu}\label{ey2}
\end{eqnarray}
resulting from the constants of motion Eqs. (\ref{eg0a}) and (\ref{eg0b}) respectively are also useful in this regard.

We now consider a family of circular orbits and take a typical circular orbit of definite radius in the equatorial plane along which the
particle spin is constant and orthogonal the plane of orbit, as the fiducial path. Deviations from this orbit can be obtained by solving
Eqs. (\ref{e5}) and (\ref{e4}) for $n^\mu$ but before embarking on that, it would be useful to reduce the number of unknowns by using
Eqs. (\ref{e6})-(\ref{ey2}). From Eq. (\ref{ey1}) we have
\begin{equation}
j^\phi=\frac{fp^t}{a^2p^\phi}j^t,\label{adi0}
\end{equation}
and from Eq. (\ref{ey2}) we have
\begin{equation}\label{adi1}
J^{r\phi}=-\frac{p^\phi}{p^t}J^{tr}.
\end{equation}
Similarly Eq. (\ref{e6}) result in 
\begin{eqnarray}
J^{tr}&=&-\frac{{\hat s}fp^t}{ap^\phi}j^t\label{adi2},\\
J^{t\phi}&=&\frac{{\hat s}}{af}j^r\label{adi3},\\
J^{\theta\phi}&=&-\frac{fp^t}{a^2p^\phi}J^{t\theta}\label{adi4}.
\end{eqnarray}
Now, taking Eqs. (\ref{adi0})-(\ref{adi4}) into account, from Eq. (\ref{e5}) with $\mu=t,r,\theta,$ and $\phi$ we obtain
\begin{equation}
\frac{d j^t}{d\tau}=-\frac{{\hat s}af^{\prime\prime}p^\phi}{2f}\frac{d n^r}{d\tau}+\frac{f^\prime}{2}(v^t-{\hat s}v^\phi)\left(\frac{f^\prime 
p^t}{2}n^t-ap^\phi n^\phi-\frac{j^r}{f}\right),\label{w1}
\end{equation}
\begin{eqnarray}
\frac{d j^r}{d\tau}&=&-\frac{{\hat s}aff^{\prime\prime}p^\phi}{2}\frac{d n^t}{d\tau}+\frac{{\hat s}ff^\prime p^t}{2}\frac{d n^\phi}
{d\tau}+\frac{{\hat s}aff^{\prime\prime\prime}v^tp^\phi}{2}n^r\nonumber\\&&\hspace{-2mm}+(v^t-{\hat s}v^\phi)\left(\frac{p^t}
{4}({f^\prime}^2-2ff^{\prime\prime})n^r+\frac{m^2ff^\prime}{2a^2{p^\phi}^2}j^t\right)\label{w2},
\end{eqnarray}
\begin{eqnarray}
\frac{d j^\theta}{d\tau}&=&\left(\frac{ff^\prime}{2a}v^t+\frac{f(1-f)}{a^2p^\phi}v^\phi p^t\right)J^{t\theta}\nonumber\\&&+\left(-v^\phi
p^\phi+\frac{{\hat s}ff^\prime}{2a}v^tp^\phi+\frac{{\hat s}f(1-f)}{a^2}v^\phi p^t\right)n^\theta,\label{w3}\\
\frac{d j^t}{d\tau}&=&-\frac{{\hat s}f^{\prime}p^\phi}{2f}\frac{d n^r}{d\tau}\nonumber\\&&+\frac{a{p^\phi}^2}{{p^t}^2}(v^t-{\hat 
s}v^\phi)\left(\frac{f^\prime p^t}{2}n^t-ap^\phi n^\phi-\frac{1}{f}j^r\right),\label{w4}
\end{eqnarray}
respectively. Similarly from Eq. (\ref{e4}) with $\mu\nu=tr,t\theta,t\phi,r\theta$, and $\theta\phi$ we obtain
\begin{equation}
-{\hat s}\frac{d j^t}{d\tau}=\frac{ap^\phi}{f}\frac{d n^r}{d\tau}+\frac{ap^\phi}{p^t}(v^t-{\hat s}v^\phi)\left(\frac{f^\prime p^t}{2}n^t-ap^\phi 
n^\phi-\frac{1}{f}j^r\right),\label{w5}
\end{equation}
\begin{equation}
\frac{d J^{t\theta}}{d\tau}=-\frac{f^\prime}{2f}v^tJ^{r\theta}-j^\theta v^t+p^t\frac{d n^\theta}{d\tau},\label{w6}
\end{equation}
\begin{eqnarray}\label{w7}
2{\hat s}\frac{d j^r}{d\tau}&=&2af(p^t\frac{d n^\phi}{d\tau}-p^\phi\frac{d n^t}{d\tau})-\frac{2m^2f(v^t-{\hat s}v^\phi)}
{ap^tp^\phi}j^t\nonumber\\&&+(2fp^tv^\phi-af^\prime p^\phi v^t+a{\hat s}f^\prime p^\phi v^\phi-{\hat s}ff^{\prime\prime}v^tp^t)n^r
\end{eqnarray}
\begin{eqnarray}
\frac{d J^{r\theta}}{d\tau}&=&-\left(\frac{ff^\prime}{2}v^t-\frac{f^2v^\phi p^t}{ap^\phi}\right)J^{t\theta}-\left(\frac{{\hat s}ff^\prime}
{2}v^t p^\phi+\frac{{\hat s}f(1-f)}{a}v^\phi p^t\right)n^\theta,\label{w8}
\end{eqnarray}
\begin{equation}\label{w9}
-\frac{fp^t}{a^2p^\phi}\frac{d J^{t\theta}}{d\tau}=\frac{1}{a}v^\phi J^{r\theta}+v^\phi j^\theta-p^\phi\frac{d n^\theta}{d\tau}
\end{equation}
respectively. Note that setting $\mu\nu=r\phi$ lead to Eq. (\ref{w5}) again. 

Eqs. (\ref{w1}), (\ref{w2}), (\ref{w4}), (\ref{w5}), and (\ref{w7}) result in $$n^r=const.$$ which is simply separation between two 
adjacent circular orbits, and $$j^t=const., j^r=const., \frac{d n^t}{d\tau}=const., \frac{d n^\phi}{d\tau}=const.$$ Thus radial perturbations do 
not lead to instability. 

\begin{figure}[h]
\vspace{5mm}
\begin{center}
\includegraphics[width=7cm]{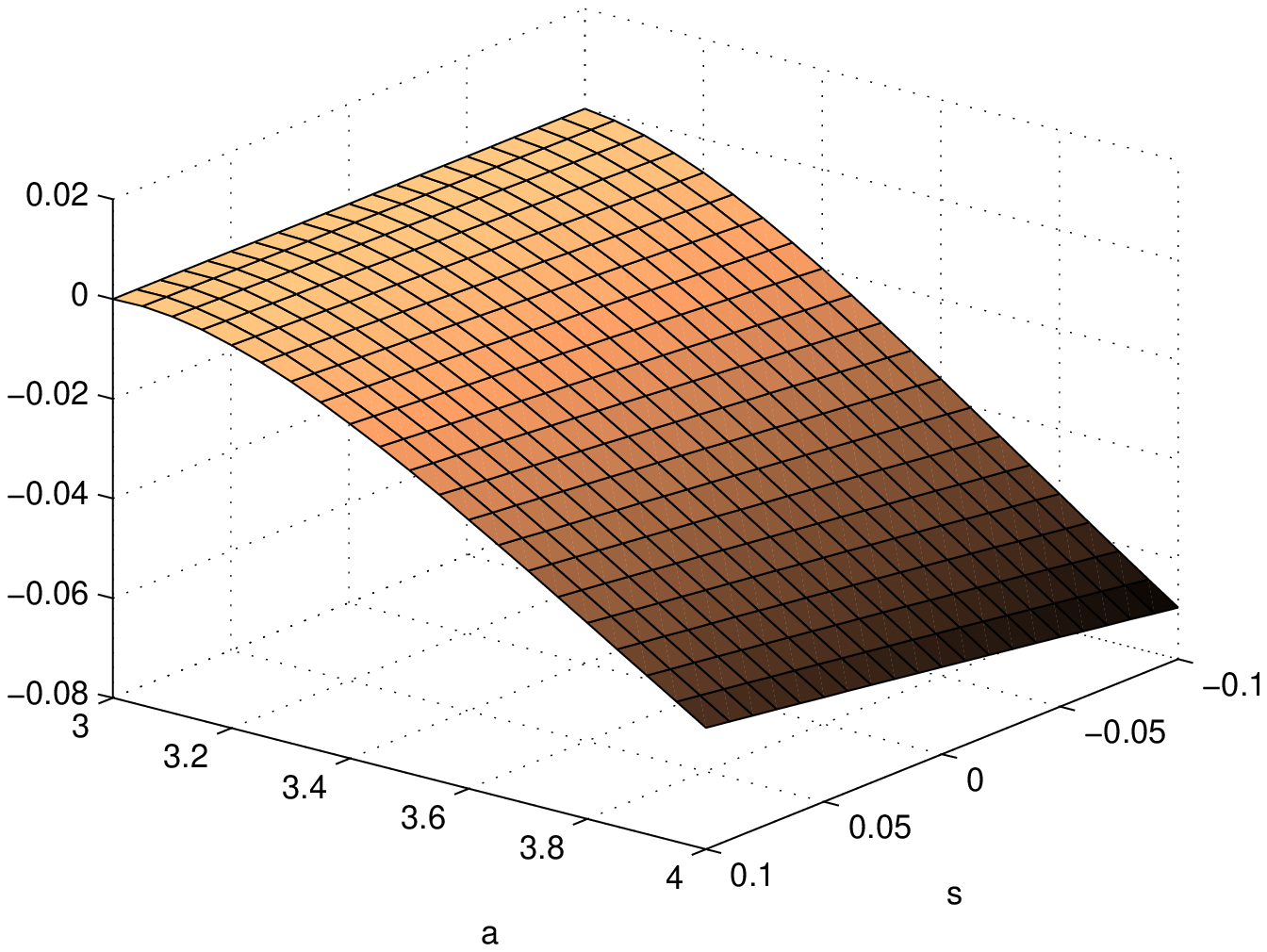}\includegraphics[width=7cm]{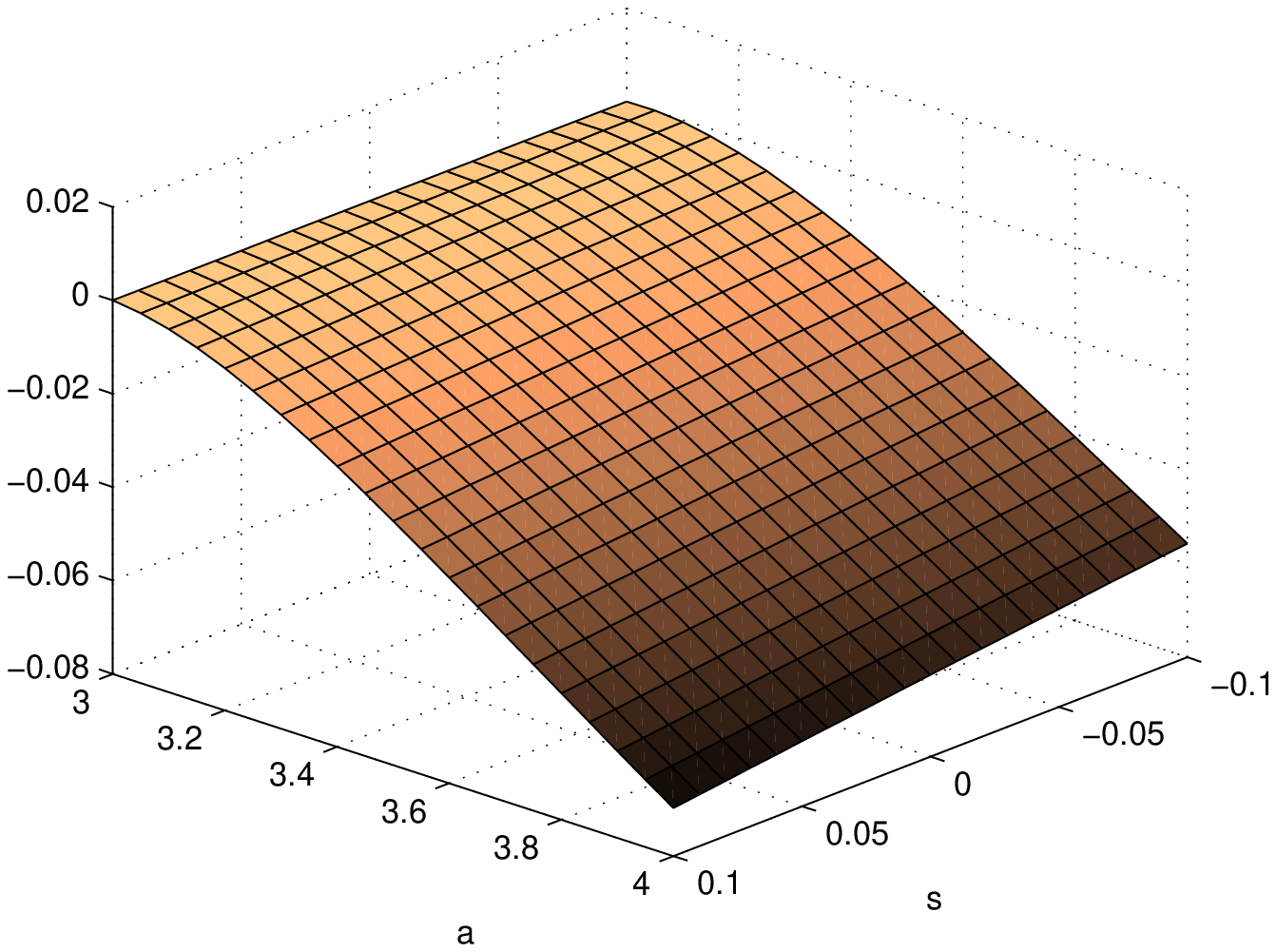}
\end{center}
\caption{$\alpha_+$ (left) and $\alpha_-$ (right) in terms of $a,{\hat s}$.}\label{fig1}
\end{figure}

The solutions to Eqs. (\ref{w3}), (\ref{w6}), (\ref{w8}), and (\ref{w9}) are given by
\begin{equation}\label{kom1}
n^\theta_\pm(\tau)=D_1\sin(v^\phi_\pm\tau)+D_2\cos(v^\phi_\pm\tau) +D_3\exp(\kappa_\pm\tau)+D_4\exp(-\kappa_\pm\tau)
\end{equation}
and similar expressions for $j^\theta(\tau), J^{t\theta}(\tau), J^{r\theta}(\tau)$. Here $D_1, D_2, D_3, D_4$ are constants, 
\begin{eqnarray*}
\kappa_\pm=\frac{\sqrt{\alpha_\pm}v^tp^t}{2am},
\end{eqnarray*}
\begin{eqnarray*}
\alpha_\pm=(2f-af^\prime)(a^3f^\prime\beta_\pm^2-2a^2f\beta_\pm\gamma_\pm+{\hat s}aff^\prime\beta_\pm+2{\hat s}f(1-f)\gamma_\pm),
\end{eqnarray*}
\begin{eqnarray*}
\beta_\pm=\frac{-({\hat s}^3{f^\prime}^2+{\hat s}a^2f^{\prime\prime}-3a{\hat s}f^\prime\pm ah)}{a
({\hat s}^2af^{\prime\prime}+{\hat s}^2f^\prime-4a\pm{\hat s}h)},
\end{eqnarray*}
\begin{eqnarray*}
\gamma_\pm=\frac{-{\hat s}f^\prime+{\hat s}af^{\prime\prime}\pm h}{2(2a-{\hat s}^2f^\prime)}.
\end{eqnarray*}
Note that under ${\hat s}\leftrightarrow{-\hat s}$ we have $\alpha_+\leftrightarrow\alpha_-$. Thus for $\alpha_\pm<0$ we have oscillatory $\sin$ 
and $\cos$ terms only and the orbits are stable. For $\alpha_\pm>0$ there are both oscillatory and exponential terms and the orbits are not in 
general stable. Depending on the values of space-time parameters and the particles specific spins ${\hat s}$, both cases may occur. For the case 
of Schwarzschild space-time we have depicted $\alpha_\pm$ in terms of $a$ and ${\hat s}$ in Fig. \ref{fig1}. According to this figure both 
solutions corresponds to stable orbits in Schwarzschild space-time, at least for the range of parameters chosen here. 
\section{Discussion}
We have studied circular orbits of spinning particles with their spins perpendicular to the plane of motion in spherically symmetric space-
times. We have shown that circular orbits of spinning particles with their velocity and momentum being parallel are only allowed in 
de Sitter case. For other cases the particle velocity and momentum are not parallel. We have obtained general expressions for the 
particle angular velocities and momenta in terms of the radii of orbits and specific spins. Depending on the sense of rotation with respect to 
the spin direction, there are in general two physically equivalent sets of such expressions for any given radius. According to these 
expressions, for a fixed spin direction and orbit, the absolute values of angular velocities of particles orbiting in opposite senses are 
different. This is due to the spin-dependent force exerted on particles. For the case of Schwarzschild space-time, there is no circular orbit 
of radius smaller than three (measured in units of the black hole mass). This is in agreement with the results of \cite{svir}. The effects due 
to spin decrease when the radius of the orbits increase because the curvature and hence the spin-curvature coupling decreases with radius.   
        
We have also solved the world-line deviation equations of spinning particles of the same specific spins moving on orbits described above. The 
resulting deviation vectors have been used to show that the orbits are stable against radial perturbations. We have determined the general 
condition under which the orbits are stable against normal perturbations. It shows that in general particles orbiting in opposite senses have 
different deviations which may result in different stability properties. By explicit calculations we have shown that the orbits are stable in 
Schwarzschild space-time. An advantage of our approach is that some rather general results can be extracted from the equations without resort to 
detailed numerical calculations. These results might be of interest in studies related to (in)stability of non-geodesic flows. The deviations 
$n^\mu$ obtained here enable us to compute the corresponding Lyapunov indices. Although the special case studied here, i.e. exactly circular 
orbits with a perpendicular spin may not be realized in real astrophysical situations, it can be regarded as a toy model which can be extended 
to more realistic situations. Possible interesting extensions include a study of stability of spinning particle orbits in a Kerr-type background 
and charged spinning particles in a Reissner-Nordstr\"{o}m space-time.

\subsection*{Acknowledgements}
I wish to acknowledge the support of the Abdus Salam ICTP, Trieste, where part of this work was done. I am also indebted to 
anonymous referees of General Relativity and Gravitation for valuable comments.  


\begin{thebibliography}{10}
\bibitem{cohen}Cohen, J.M., Mashhoon, B.: Phys. Lett. A {\bf 181}, 353 (1993)
\bibitem{pringle}Pringle, J.E.: Ann. Rev. Astron. Astrophys. {\bf 19}, 137 (1981)
\bibitem{kato}Kato, S., Fukue, J., Mineshige, S.: Black-hole Accretion Disks, Kyoto University Press, Kyoto (1998) 
\bibitem{mathi}Mathisson, M.: Acta. Phys. Pol. {\bf 6}, 163 (1937)
\bibitem{papa}Papapetrou, A.: Proc. Roy. Soc. Lond. A {\bf 209}, 248 (1951) 
\bibitem{tulc}Tulczyjew, W.: Acta. Phys. Pol. {\bf 18}, 393 (1959) 
\bibitem{dixon}Dixon, W.G.: Proc. Roy. Soc. Lond. A {\bf 314}, 499 (1970) 
\bibitem{tod}Tod, K.P., de Felice, F., Calvani, M.: Nuovo Cim. B {\bf 34}, 365 (1976)
\bibitem{ryab1}Ryabushko, A.P., Bakhankov, A.A.: Gen. Rel. Grav. {\bf 19}, 351 (1987)
\bibitem{svir} Svirskas, K., Pyragas, K., Lozdien{\'e}, A.: Astrophys. Space Sci. {\bf 149}, 39 (1988)
\bibitem{ryab2}Bakhankov, A.A., Ryabushko, A.P.: Ann. Phys. (Leipzig) {\bf 501}, 605 (1989)
\bibitem{bini}Bini, D., de Felice, F. Geralico, A.: Class. Quantum Grav. {\bf 21}, 5427 (2004)
\bibitem{geralico2}Bini, D., de Felice, F., Geralico, A., Jantzen. R.T.: Class. Quantum Grav. {\bf 22}, 2947 (2005) 
\bibitem{plya}Plyatsko, R.: Class. Quantum Grav. {\bf 22}, 1545 (2005)
\bibitem{bakh}Bakhankov, A.A., Ryabushko, A.P.: Gen. Rel. Grav. {\bf 21}, 447 (1989)
\bibitem{seme}Semer{\'a}k, O.: Mon. Not. R. Astron. Soc. {\bf 308}, 863 (1999)
\bibitem{hart1}Hartl, M.D.: Phys. Rev. D {\bf 67}, 024005 (2003) 
\bibitem{hart2}Hartl, M.D.: Phys. Rev. D {\bf 67}, 104023 (2003)
\bibitem{faru}Faruque, S.B.: Phys. Lett. A {\bf 327}, 95 (2004)
\bibitem{geralico}Bini, D., de Felice, F., Geralico, A.: Class. Quantum Grav. {\bf 21}, 5441 (2004)
\bibitem{felice}Bini, D., de Felice, F., Geralico, A., Jantzen, R.T.: Class. Quantum Grav. {\bf 23}, 3287 (2006)
\bibitem{singh}Mashhoon, B., Singh. D.: Phys. Rev. D {\bf 74}, 124006 (2006)
\bibitem{kyr}Kyrian, K., Semer{\'a}k, O.: Mon. Not. R. Astron. Soc. {\bf 382}, 1922 (2007)
\bibitem{singh2}Singh. D.: Phys. Rev. D {\bf 78}, 104028 (2008)
\bibitem{bini2}Bini, D., de Felice, F., Geralico, A.: Int. J. Mod. Phys. D {\bf 14}, 1793 (2005)
\bibitem{binib}Bini, D., De Felice, F., Geralico, A., Lunari, A.: J. Phys. A {\bf 38}, 1163 (2005) 
\bibitem{van}Rietdijk, R.H., van Holten, J.W.: Class. Quantum Grav. {\bf 10}, 575 (1993)
\bibitem{rit}van Holten, J.W., Rietdijk, R.H., J. Geom. Phys. {\bf 11}, 559 (1993) 
\bibitem{gib}Gibbons, G.W., Rietdijk, R.H., van Holten, J.W.: Nucl. Phys. B {\bf 404}, 42 (1993)
\bibitem{shir} Shirokov, M.F.: Gen. Rel. Grav. {\bf 4}, 131 (1973)
\bibitem{fuchs}Fuchs, H.: Astron. Nachr. {\bf 311}, 271 (1990)
\bibitem{rosa}Rosa, V.M., Letelier, P.S.: Phys. Rev. D {\bf 78}, 084038 (2008)
\bibitem{wu}Wu, X., Huang, T-Y., Zhang, H. Phys. Rev. D {\bf 74}, 083001 (2006) 
\bibitem{suzuki}Suzuki, S., Maeda, K.: Phys. Rev. D {\bf 55}, 4848 (1997)
\bibitem{sano} Kao, J-K., Cho, H.T.: Phys. Lett. A {\bf 336}, 159 (2005) 
\bibitem{suz} Suzuki, S., Maeda, K.: Phys. Rev. D {\bf 58}, 023005 (1998)
\bibitem{nieto}Nieto J.A., Saucedo, J., Villanueva, V.M.: Phys. Lett. A {\bf 312}, 175 (2003)
\bibitem{mohseni}Mohseni, M.: Phys. Lett. B {\bf 587}, 133 (2004)
\bibitem{sepangi}Heydari-Fard, M., Mohseni, M., Sepangi, H.R.: Phys. Lett. B {\bf 626}, 230 (2005)
\bibitem{muller}Moeller, C.: Commun. Dublin Inst. Adv. Studies A {\bf 5}, 3 (1949)
\bibitem{ehler}Ehlers, J., Rudolph, E.: Gen. Rel. Grav. {\bf 8}, 197 (1977) 
\end{thebibliography}
\end{document}